\documentclass[12pt]{article}
\usepackage{graphicx}
\usepackage{amsmath,amssymb}
\usepackage{amsfonts}
\usepackage{array}
\usepackage{lineno}
\newsavebox{\anglearrow}
\savebox{\anglearrow}{%
\begin{picture}(10,15)(0,0)
\put(0,15){\line(0,-1){10}}
\put(0,5){\vector(1,0){10}}
\end{picture}}
\newcommand\pubnumber{}
\newcommand\pubdate{\today}

\def\rcns{RCNS, Tohoku Univ. Sendai, 980-8578, JAPAN \\
Faculty of Science, Toho Univ., Chiba, 274-8510, JAPAN}
\def\support{\footnote{Supported by the JSPS.}}

\def\Title#1{\begin{center} {\Large #1 } \end{center}}
\def\Author#1{\begin{center}{ \sc #1} \end{center}}
\def\Address#1{\begin{center}{ \it #1} \end{center}}

\newcommand\pubblock{\rightline{\begin{tabular}{l} \pubnumber\\
         \pubdate  \end{tabular}}}
\newenvironment{Abstract}{\begin{quotation}  }{\end{quotation}}
\newenvironment{Presented}{\begin{quotation} \begin{center} 
             PRESENTED AT\end{center}\bigskip 
      \begin{center}\begin{large}}{\end{large}\end{center} \end{quotation}}

\def\beq{\begin{equation}}
\def\eeq#1{\label{#1}\end{equation}}
\def\eeqn{\end{equation}}

\def\beqa{\begin{eqnarray}}
\def\eeqa#1{\label{#1}\end{eqnarray}}
\def\eeqan{\end{eqnarray}}

\let\bar=\overbar

\def\Dslash{\not{\hbox{\kern-4pt $D$}}}
\def\dslash{\not{\hbox{\kern-2pt $\del$}}}

\def\msb{{\bar{\ssstyle M \kern -1pt S}}}

\begin{document}
\begin{titlepage}
\pubblock

\vfill
\Title{Hunt for Sterile Neutrinos: Decay at Rest Experiments}
\vfill
\Author{Fumihiko Suekane\support}
\Address{\rcns}
\vfill
\begin{Abstract}
In the standard model of the elementary particles, the number of neutrino flavor is three.  
However, there have been indications of existence of 4th neutrino, called sterile neutrino, in some neutrino oscillation related experiments.
A number of experiments are planned to test whether such indications are true or not. 
Among them, experiments which use neutrinos from $\pi^+$, K$^+$, $\mu^+$ decay at rest (DAR) are promising because the energy spectra of neutrinos are very well known and clean oscillation measurements are possible. 
In this proceedings, properties of such DAR neutrinos and LSND, JSNS$^2$, OscSNS and KPipe experiments are briefly introduced.
\end{Abstract}
\vfill
\begin{Presented}
NuPhys2015, Prospects in Neutrino Physics\\
Barbican Centre, London, UK,  December 18, 2015 \\
\end{Presented}
\vfill
\end{titlepage}
\def\thefootnote{\fnsymbol{footnote}}
\setcounter{footnote}{0}

\section{Introduction}
\begin{figure}[htbp]
\centering
\includegraphics[width=120mm]{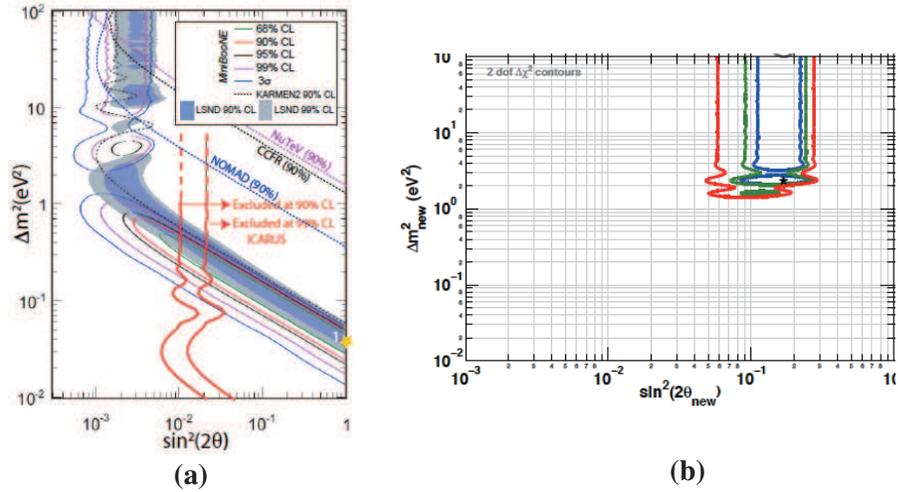}
\caption{\small{Sterile neutrino positive parameter regions. 
(a) Appearance mode~\cite{ICARUS12} (b) Disappearance mode~\cite{White12} }}
\label{fig:SterileParameters}
\end{figure}

In the standard model of the elementary particles, the number of neutrino flavors is three. 
However, some of neutrino oscillation related experiments show anomalies which can not be explained by the standard three flavor neutrino oscillations.  
LSND (Liquid Scintillator Neutrino Detector) group reported an excess of 88  $\bar{\nu}_e$ events in $\bar{\nu}_\mu$ beam in 1993$\sim$1998, where $\bar{\nu}_\mu$ was produced in $\mu^+$ decay at rest(DAR)~\cite{LSND01}. 
If the excess is caused by neutrino oscillation ($\bar{\nu}_\mu \to \bar{\nu}_e$), the mass-square difference is $\Delta m^2 > 10^{-2}~{\rm eV^2}$.
This large $\Delta m^2$ can not be explained by the standard three flavor neutrino oscillations and existence of 4th neutrino which does not feel the weak interactions and thus called {\it sterile neutrino}, has been suggested.
Later on, 
KARMEN (KArlsruhe Rutherford Medium Energy Neutrino experiment) group performed similar measurement and obtained null result~\cite{KARMEN98}. 
However, some of the LSND-positive oscillation parameter regions have survived because of shorter baseline. 
MiniBooNE (Mini Booster Neutrino Experiment) group performed measurement using neutrinos from decay in flight and 
obtained positive oscillation results~\cite{MINIBOONE13}. 
However, it is pointed out that there is a possibility that the observed $\bar{\nu}_e$ and $\nu_e$ signals are actually caused by $\gamma$'s from neutral current interactions and the result is not conclusive. 
ICARUS(Imaging Cosmic And Rare Underground Signals) group obtained negative result and showed $\sin^22\theta<10^{-2}$~\cite{ICARUS12}. 
Fig.-\ref{fig:SterileParameters}(a) shows the oscillation parameters of the above appearance experiments. 

On the other hand, reactor neutrino flux has been reported to be 6\% smaller than expectation and the $\nu_e$ flux from radioactive source has been reported to be 14\% smaller than expectation (see summary \cite{White12}). 
Those results can be explained if there are sterile neutrinos with $m>1~$eV which mix with our regular neutrinos with mixing angle $\sin^22\theta \sim 0.2$ as shown in Fig.-\ref{fig:SterileParameters}(b). 
In order to test those indications a number of experiments are being performed and planned.
This proceedings describe the properties of DAR neutrinos and LSND experiment which observed positive DAR $\bar{\nu}_\mu \to \bar{\nu}_e$ oscillation and then new generation DAR experiments, JSNS$^2$, OscSNS and KPipe.

\section{Neutrinos from $\mu^+$, $\pi^+$ and K$^+$ decay at rest}
Currently there are two planned sites for DAR experiments, one is J-PARC Material Life Science Facility (MLF) in Japan and the other is Oak Ridge Spallation Neutron Source (SNS) in the Unites States. 
In this section, properties of neutrinos from DAR are described referencing the MLF beam as an example.  

\begin{figure}[htbp]
\centering
\includegraphics[width=140mm]{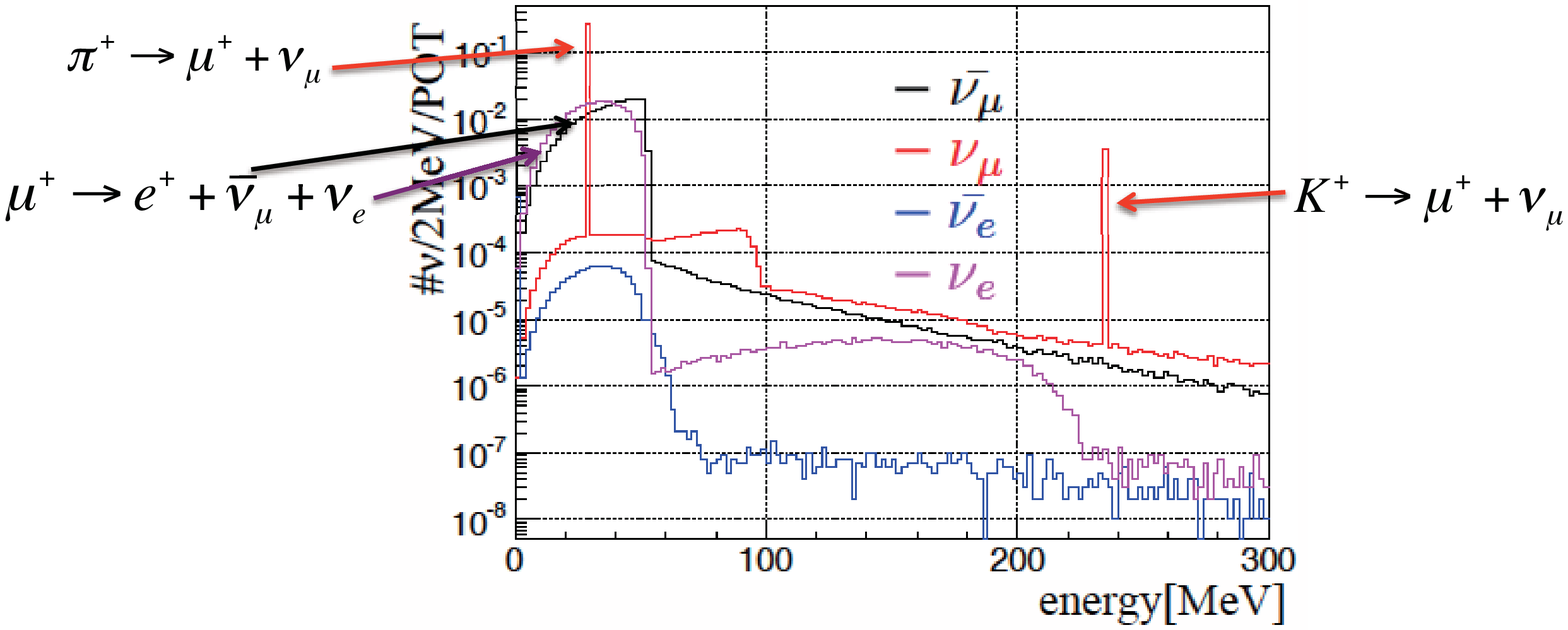}
\caption{\small{Example of the neutrino energy spectra (J-PARC MLF)~\cite{JSNS2}. The neutrinos are produced in decay at rest (DAR), decay in flight (DIF) and nuclear interactions. 
Four kinds of neutrinos from DAR to be used in the experiments described in this paper are indicated by arrows.}}
\label{fig:DARnu}
\end{figure}

The energy spectrum of the neutrinos from the MLF target is shown in 
Fig.-\ref{fig:DARnu}.
Four kinds of neutrinos from DAR to be used in the sterile neutrino experiments are indicated by arrows.
The production scheme of those DAR neutrinos is shown below. 
\begin{equation}
 \begin{split}
 &p + {\rm target} \to \pi^+/{\rm K}^+ + X \\
 & ~~~~~~~~~~~~~~~~~~~~\usebox{\anglearrow }~\pi^+/{\rm K}^+ ({\rm stop}) 
 \xrightarrow{\tau=26/12{\rm ns}} \mu^+ + \nu_\mu (E_\nu=30/236~{\rm MeV})  \\
 &~~~~~~~~~~~~~~~~~~~~~~~~~~~~~~~~~~~~~~~~~~~~~~~~~~~~~~~\usebox{\anglearrow }
  ~\mu^+({\rm stop})  \xrightarrow{\tau=2.2\mu{\rm s}} e^+ + \bar{\nu}_\mu + \nu_e~~ 
 \end{split}
 \label{eq:DAR_Process}
\end{equation}
High energy protons hit target material and produce $\pi$ and K mesons by the strong interactions. 
The $\pi^+/{\rm K}^+$ stop in the target and decay to $\mu^+ + \nu_\mu$ with lifetimes $\tau=26/12$~ns. 
Since the $\pi^+/{\rm K}^+$ decay at rest, the produced $\nu_\mu$'s have monochromatic energies of $E_\nu = 30/236~$MeV. 
The produced $\mu^+$'s stop in the target and decay as 
$\mu^+ \to e^+ + \nu_e + \bar{\nu}_\mu$ with lifetime 2.2~$\mu$s. 
Since the momentum of the parent $\mu^+$ is zero, the energy spectra of the produced $\bar{\nu}_\mu$ and $\nu_e$ are well known. 
By setting the timing window of the event selection to be less than a few tens of nano seconds from the beam pulse, the monochromatic $\nu_\mu$'s from the $\pi^+$ and K$^+$ DAR can be selected and by setting the timing window to be later than a few hundreds nano seconds, the $\bar{\nu}_\mu$ and 
$\nu_e$ from the $\mu^+$ DAR can be selected. 
For $\bar{\nu}_\mu \to \bar{\nu}_e$ oscillation experiments, there is an intrinsic background of $\bar{\nu}_e$ from the $\pi^-({\rm stop}) \to \mu^-({\rm stop}) \to \bar{\nu}_e$ decays. 
However the magnitude of the $\bar{\nu}_e$ flux is suppressed to an order of $10^{-3}$ since $\pi^-$ and $\mu^-$ are absorbed by the target nuclei before they decay.
In addition, the small contribution of the background $\bar{\nu}_e$ can be measured from the energy spectrum.

\section{Decay at rest experiments}

\subsection{LSND experiment}
\begin{figure}[htbp]
\centering
\includegraphics[width=100mm]{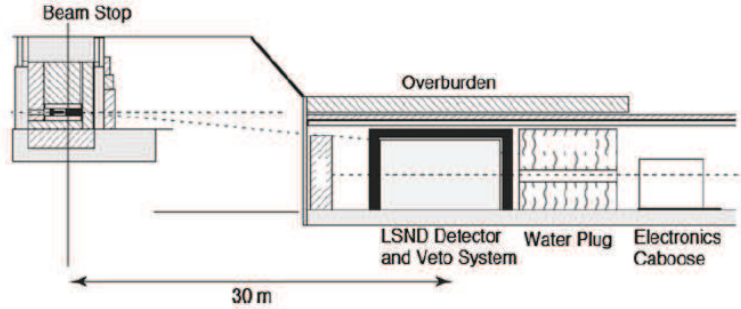}
\caption{\small{LSND experiment~\cite{LSND01}}}
\label{fig:LSND_Detector}
\end{figure}

LSND experiment was performed in 1993$\sim$1998 at Los Alamos National Laboratory and showed positive 
$\bar{\nu}_\mu \to \bar{\nu}_e$ oscillation at $\Delta m^2 > 10^{-2}~{\rm eV^2}$.
Fig.-\ref{fig:LSND_Detector} shows a side view of the LSND experiment.
The LSND experiment used LAMPF proton beam whose energy and current were 800~MeV and 1~mA, respectively.
600~$\mu s$ of pulse beam was delivered with repetition rate of 120~Hz, resulting in the duty factor of 7.2~\%.
The beam target was water or high-Z material and the beam stopper was copper. 
The baseline was 30~m. 
\begin{figure}[htbp]
\centering
\includegraphics[width=120mm]{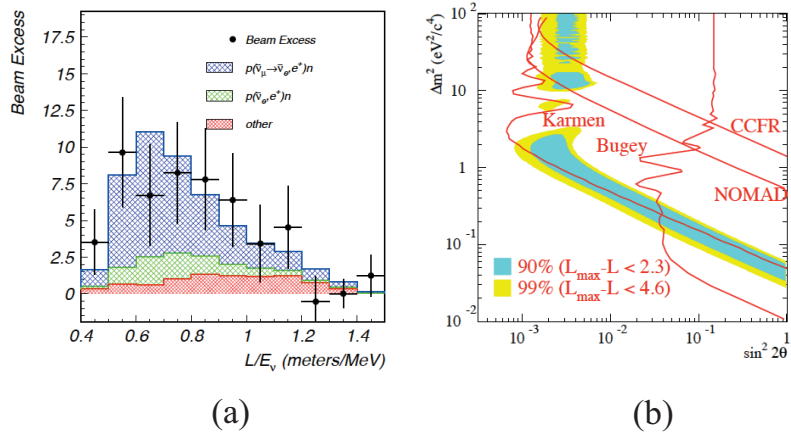}
\caption{\small{LSND result~\cite{LSND01} 
(a) Excess of $\bar{\nu}_e$ events (b) Allowed oscillation parameter regions }}
\label{fig:LSND_Result}
\end{figure}
It used a 167~tons of low light-output liquid scintillator to detect both 
\v{C}erenkov light and scintillation light.
Fast neutron background can be reduced by requiring the \v{C}erenkov light.
$\bar{\nu}_e$ was detected by inverse $\beta$ decay reaction with proton followed by the neutron capture on proton.
$$\bar{\nu}_e + p \to e^+ + n~: ~n+ p \to d + \gamma ( 2.2~{\rm MeV})$$
The average time difference between the positron signal and neutron signal was $\sim200~{\rm \mu s}$. 
They observed excess of 88 $\bar{\nu}_e$ events in 6~years operation as shown in Fig.-\ref{fig:LSND_Result}(a).  
The allowed oscillation parameter regions are shown in Fig.-\ref{fig:LSND_Result}(b).
This positive oscillation result is inconsistent with other neutrino oscillation measurements within the three neutrino flavor scheme and has not been accepted as an conclusive result from the neutrino oscillation community and further experiments with better sensitivities are required to test the result.

\subsection{JSNS$^2$ experiment}
\begin{figure}[htbp]
\centering
\includegraphics[width=120mm]{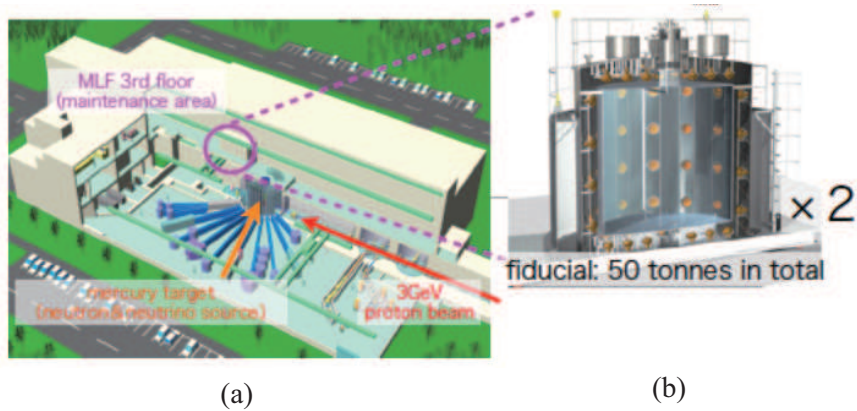}
\caption{\small{JSNS$^2$ experiment~\cite{JSNS2} (a) experimental site (b) detector}}
\label{fig:JSNS2exp}
\end{figure}
\begin{figure}[htbp]
\centering
\includegraphics[width=120mm]{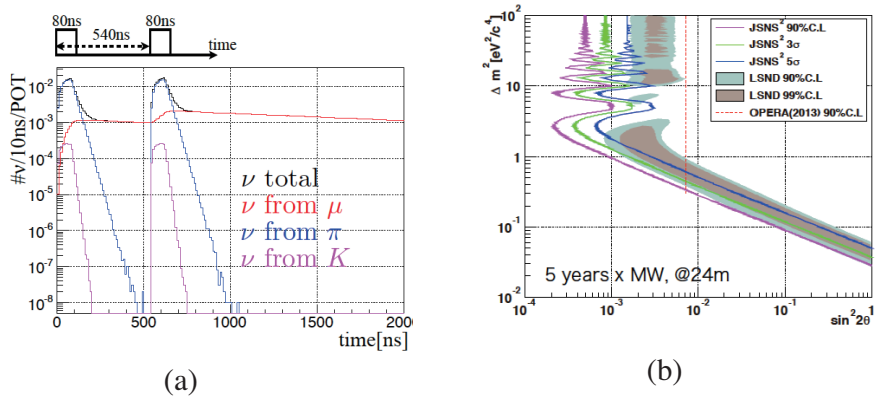}
\caption{\small{~ (a) Timing spectrum and (b) Expected Sensitivity of JSNS$^2$\cite{JSNS2} }}
\label{fig:DAR_Nu_Spectra}
\end{figure}

JSNS$^2$(J-PARC Sterile Neutrino Search at J-PARC Spallation Neutrino Source) experiment~\cite{JSNS2} uses the DAR $\bar{\nu}_\mu$ from the J-PARC MLF beam line shown in Fig.-\ref{fig:JSNS2exp}(a).
The energy of the proton beam is 3~GeV and the power will become 1~MW when the experiment is supposed to starts. 
The MLF proton beam consists of two narrow ($\sim$100~ns) pulses which are 
$\sim$600~ns apart. 
The twin beams hit the target every 40~ms (25~Hz).
Fig.-\ref{fig:DAR_Nu_Spectra}(a) shows the timing of the neutrino production.
By setting the timing window $(1<t<10~{\rm \mu s})$ after the start of the first beam pulse, the beam associate background and neutrinos from $\pi$/K decays can be eliminated and the beam uncorrelated background can be suppressed to $1.1\times 10^{-4}$.
The $\bar{\nu}_e$ background from the $\pi^-({\rm stop}) \to \mu^-({\rm stop}) \to \bar{\nu}_e$ is suppressed to $1.7\times 10^{-3}$ because $\pi^-$ and $\mu^-$ are absorbed by high-Z (mercury) target nuclei.

JSNS$^2$ will use Gadolinium loaded liquid scintillator (Gd-LS) as neutrino target. 
Two neutrino detectors containing 25~tons Gd-LS each (Fig.-\ref{fig:JSNS2exp}(b)) will be used. 
The detection method is the inverse $\beta$ decay, like LSND.  
However, the neutron is absorbed by Gd, emitting 8~MeV $\gamma$-rays. 
%
%
$$\bar{\nu}_e + p \to e^+ + n~:~n+ {\rm Gd} \to {\rm Gd'^*}~:~ 
{\rm Gd'^*} \to {\rm Gd'} + \gamma's (\Sigma E_\gamma = 8~{\rm MeV})
$$
Using Gd, the environmental $\gamma$-ray backgrounds can be eliminated and the coincidence window can be made narrower, 
($200~{\rm \mu s } \to 30~{\rm \mu s }$). 
There are two options for the liquid scintillator (LS). 
One is high light output LS with enhanced pulse shape discrimination (PSD) capability for the fast neutron rejection and the other is LSND type low light output LS. 
The baseline is 24~m and 
the number of events will be 100/year in case $\sin^22\theta = 0.003 $ and $\Delta m^2 > 1~{\rm eV}^2$.
%
The neutrino production and detection mechanisms are the same as those of LSND's and a direct test of LSND result can be performed.  
The sensitivity of JSNS$^2$ is shown in Fig.-\ref{fig:DAR_Nu_Spectra}(b).

The JSNS$^2$ group submitted the proposal in 2013 and performed on-site background measurements using 500~kg plastic scintillators and obtained stage-1 approval from J-PARC PAC in 2014. 
The group is now requesting the budget for construction of the detector. 

\subsection{OscSNS experiment}

\begin{figure}[htbp]
\centering
\includegraphics[width=120mm]{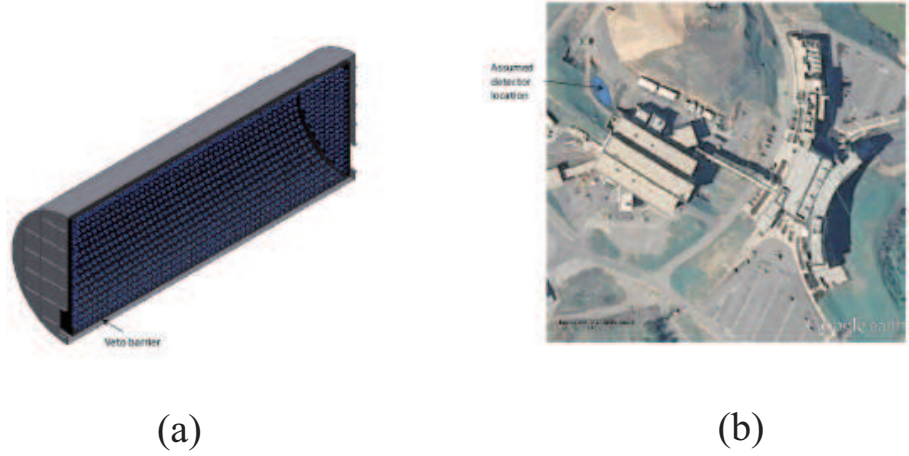}
\caption{\small{OscSNS experiment~\cite{OscSNS} (a) The detector (b) Oak Ridge SNS and a candidate detector location}}
\label{fig:OscSNSexp}
\end{figure}

OscSNS group is proposing to perform sterile neutrino experiment using 
the Oak Ridge Spallation Neutron Source (SNS)~\cite{OscSNS}. 
The neutrino detector and expected location are shown in 
Figs.-\ref{fig:OscSNSexp}. 
The energy of the beam is 1~GeV and power is 1.4~MW. 
The width of the beam pulse is 500~ns and the frequency is 60~Hz. 
Therefore, by setting the timing window $(1<t<10~{\rm \mu s} )$,
the beam un-correlated background can be suppressed to $2.6\times 10^{-4}$. 
The baseline to the detector is 50~m and the target mass is 450~ton. 
The liquid scintillator is LSND type to detect both \v{C}erenkov and scintillation lights. 
\begin{figure}[htbp]
\centering
\includegraphics[width=120mm]{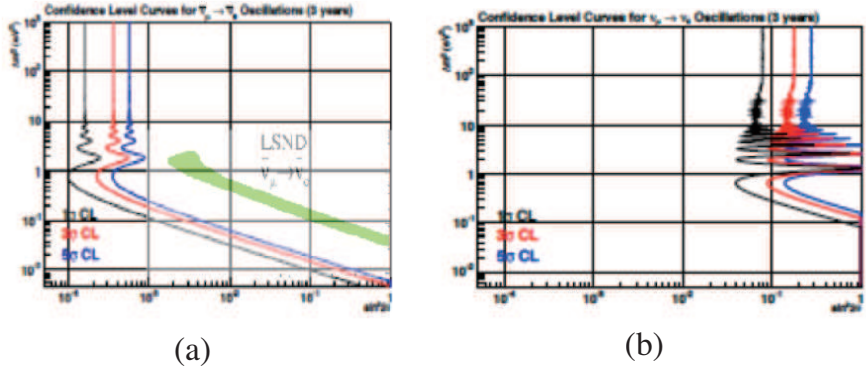}
\caption{\small{(a) Sensitivity for $\bar{\nu}_\mu \to \bar{\nu}_e$ appearance search 
(b) Sensitivity for $\nu_\mu \to \nu_\mu$ disappearance search~\cite{OscSNS} }}
\label{fig:OscSNS_Sensitivity}
\end{figure}
%
Fig.-\ref{fig:OscSNS_Sensitivity}(a) shows the sensitivity of the OscSNS for $\bar{\nu}_\mu \to \bar{\nu}_e$ measurement. 
Since the baseline is much longer than LSND, the positive region of the LSND result can be completely covered. 

Table \ref{tab:Comparison} compares main features of $\bar{\nu}_\mu \to \bar{\nu}_e$ appearance measurements of JSNS$^2$, OscSNS and LSND.
This shows both JSNS$^2$ and OscSNS are expected to have much better sensitivities than LSND experiment.


\begin{table}[htbp]
\small{
\begin{center}
\begin{tabular}{|l||c|c|c|}  
\hline
     &  JSNS$^2$ &  OscSNS &  LSND \\ 
\hline \hline
 Accelerator      & J-PARC MLF  & Oak Ridge SNS      & LAMPF \\
\hline
 Beam Energy (Power)   &  3GeV (1MW)    &   1GeV (1.4MW)      &  0.8GeV(0.8MW) \\ 
 \hline
  BKG suppression & $1.1\times 10^{-4}$   &  $2.6\times 10^{-4}$  &  0.072 \\
  by pulse beam  &                       &                       & \\
  \hline
     Liquid Scintillator & 50~t (Gd Loaded) & 450~t(LSND type)  & 167~t \\
 \hline
 Baseline  &  24~m    &   50~m      &  30~m \\ 
 \hline
  \# of $\bar{\nu}_e$(BKG) events  & 500(300)/5yr  &  600(200)/5yr  &   88/6yr$^{a)}$  \\ 
 \hline
 Stopping $\mu^-/\mu^+$&$1.7\times 10^{-3}$&$\sim 10^{-3}$& $6.5\times 10^{-4}$ \\ 
  \hline
  Delayed Coin.: (E,$\Delta t$) &  (8MeV,$30\mu$s)  &  (2.2MeV, $200\mu$s) &  (2.2MeV, $200\mu$s) \\ 
  \hline
  Fast $n$ rejection & PSD$^{b)}$ or \v{C}erenkov$^{c)}$ &  \v{C}erenkov   & \v{C}erenkov \\ 
 \hline
   $\Delta E/ E$ & 3\% @ 35MeV$^{b)}$ & - & 7\% @ 45MeV \\
  \hline
     Cost & \$ & \$\$ & - \\
  \hline
\end{tabular}
\caption{\small{Comparison of baseline designs of JSNS$^2$, OscSNS and LSND for $\bar{\nu}_\mu \to \bar{\nu}_e$ search. $\sin^22\theta=0.003$ is assumed for JSNS$^2$ and $P(\bar{\nu}_\mu \to \bar{\nu}_e) =0.26\%$ is assumed for OscSNS. 
 $^{a)}$~after very tight cuts. $^{b)}$~for high light output LS option, $^{c)}$ for low light output LS option.}}
\label{tab:Comparison}
\end{center}
}
\end{table}

In addition to the detection of $\bar{\nu}_\mu \to \bar{\nu}_e$ oscillation, OscSNS plans to detect $\nu_\mu \to \nu_e$ oscillation with following process. 
\begin{equation}
 \begin{split}
 &\pi^+({\rm stop}) \to \nu_\mu (30~{\rm MeV}) + \mu^+ \\
 & ~~~~~~~~~~~~~~~~~~\usebox{\anglearrow}~ 
 \nu_\mu \xrightarrow{\rm oscillation} \nu_e (30~{\rm MeV}) \\
 & ~~~~~~~~~~~~~~~~~~~~~~~~~~~~~~~~~~~~~~\usebox{\anglearrow}~ 
 \nu_e + ~^{12}{\rm C} \to ~^{12}{\rm N_{gs}}({\rm 17.3~MeV}) + e^-(12.5~{\rm MeV})  \\
 &~~~~~~~~~~~~~~~~~~~~~~~~~~~~~~~~~~~~~~~~~~~~~~~~~~~~~~~~~~~~~~~\usebox{\anglearrow}~
 ^{12}{\rm N_{gs}} \xrightarrow{\rm \tau=11ms}  ~^{12}{\rm C} + e^+ + \nu_e
 \end{split}
\end{equation}
The $\nu_e$ can be identified by the delayed coincidence of 12.5{\rm MeV} monochromatic electron and $\beta^+$( Q=17~MeV) signals. 

Since the detector is long, disappearance of the neutrinos due to oscillation can be measured through $L$ dependence of the deficit.
$$ \nu_x + ~^{12}{\rm C} \to \nu_x + ~^{12}{\rm C^*}(15~{\rm MeV})$$ 
$$~~~~~~~~~~~~~~~~~\nu_e + ~^{12}{\rm C} \to e^- + ~^{12}{\rm N_{gs}} \xrightarrow{\rm 11~ms} ~^{12}{\rm C} + e^+ +\nu_e$$.
Fig.-\ref{fig:OscSNS_Sensitivity}(b) shows the $\nu_\mu \to \nu_\mu$ sensitivity. 
\subsection{KPipe experiment}
Because the beam energy is high (3~GeV), monochromatic (236~MeV) $\nu_\mu$ from K$^+$ DAR are abundant at J-PARC MLF as can be seen in Fig.-\ref{fig:DARnu}.
The energy of the $\nu_\mu$ is high enough to perform the charged current interactions,
\begin{equation}
 \nu_\mu (236~{\rm MeV}) + ~^{12}{\rm C} \to \mu^- + {\rm X}.
 \label{eq:NuMu_Charged Current}
\end{equation}
Therefore, the $\nu_\mu$ can be identified by the existence of the $\mu^-$ in the final state. 
If oscillation $\nu_\mu \to \nu_S$ takes place, it is identified as a deficit of $\nu_\mu$ flux. 
Since the neutrino energy is unique, a clear oscillation pattern will be observed in the $L$ dependence of the deficit. 
The analysis is rather simple because it is not necessary to know absolute neutrino flux nor detection efficiency. 

\begin{figure}[htbp]
\centering
\includegraphics[width=110mm]{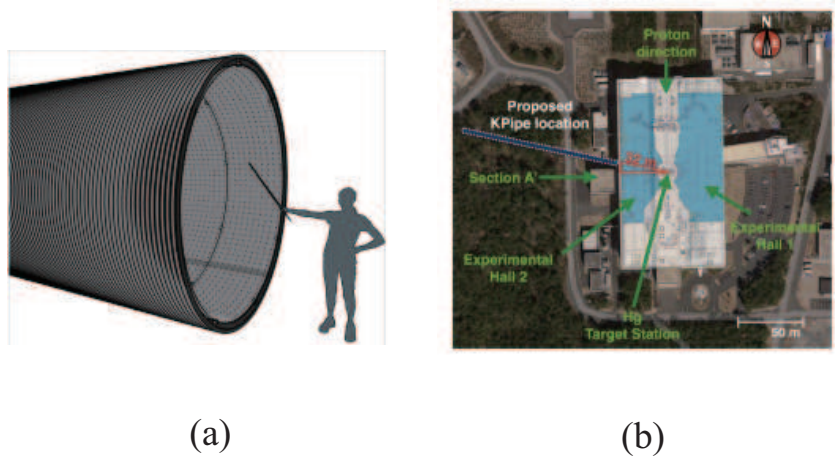}
\caption{\small{(a) KPipe detector (b) KPipe site~\cite{KPipe}}}
\label{fig:KPipe}
\end{figure}

\begin{figure}[htbp]
\centering
\includegraphics[width=140mm]{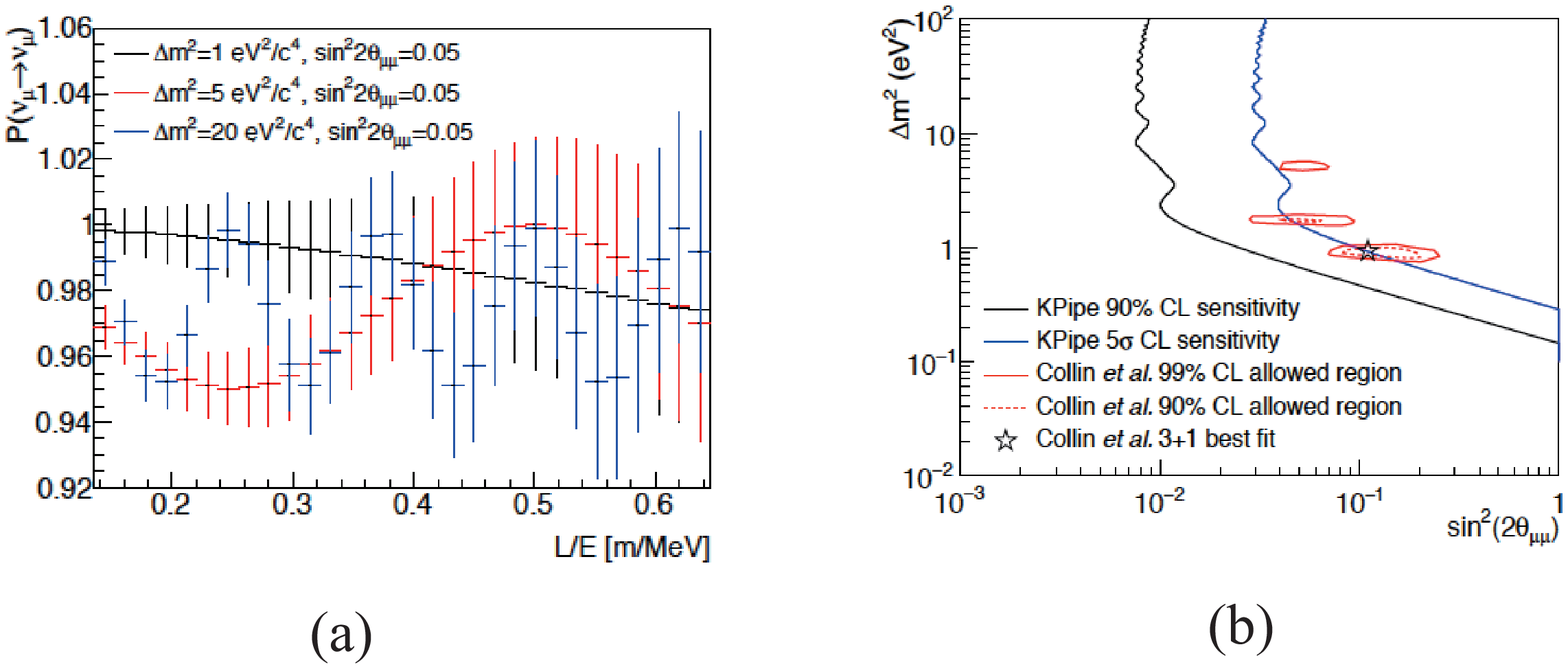}
\caption{\small{(a) $L/E$ dependence of the deficit (b) KPipe Sensitivity~\cite{KPipe}}}
\label{fig:KPipeBeamSensitivity}
\end{figure}

KPipe group is proposing to install 120~m long cylindrical liquid scintillator detector near the J-PARC MLF beam line as shown in Figs.-\ref{fig:KPipe}~\cite{KPipe}. 
%
%
The diameter of the "pipe" is 3~m and it extends 32$\sim$152~m from the target.
The total mass of the liquid scintillator is 700 ton. 
Fig.-\ref{fig:KPipeBeamSensitivity}(a) shows the $L/E$ dependence of the relative $\nu_\mu$ rate for various $\Delta m^2$ parameters and 
Fig.-\ref{fig:KPipeBeamSensitivity}(b) shows the sensitivity of this experiment.  
%
\section{Summary}
Neutrinos from $\pi^+$, $\mu^+$ and K$^+$ decays at rest are strong tool to search for the sterile neutrinos. 
LSND showed positive results sometime ago and currently JSNS$^2$, OscSNS and KPipe are proposed 
to search for the sterile neutrinos with better sensitivities than the LSND experiment, all using the DAR neutrinos.

\end{document}